\documentclass[12pt]{article}
\usepackage{amsmath}
\usepackage{graphicx}
\usepackage{enumerate}
\usepackage{natbib}
\usepackage{url} 

\newcommand{\blind}{0}

\usepackage{amssymb}
\usepackage{scalefnt}
\usepackage{amsfonts}
\usepackage{algorithmicx}
\usepackage{algorithm,algpseudocode}
\usepackage{setspace}
\usepackage{breqn}
\usepackage{microtype}      
\usepackage{graphicx}
\usepackage{subcaption}
\usepackage{tabularx}
\usepackage{amssymb,amsthm}

\usepackage[font=small,labelfont=bf]{caption}

\usepackage{blindtext,mathtools,authblk}
\usepackage{multirow,float}
\usepackage{anyfontsize}
\usepackage{booktabs,geometry}
\usepackage{tabularx,array}
\usepackage{makeidx,url}
\usepackage{color,bbm,bm}

\usepackage{comment}

\newcommand{\beqa}{\begin{eqnarray}}
\newcommand{\eeqa}{\end{eqnarray}}

\usepackage{pgfpages}
\usepackage{tikz}
\usetikzlibrary{shapes.geometric, arrows, calc, fit, positioning}
\tikzset{
    between/.style args={#1 and #2}{
         at = ($(#1)!0.5!(#2)$)
    },
    node distance=2cm
}
\definecolor{deepskyblue}{rgb}{0, 0.75, 1}
\definecolor{stateblue}{rgb}{0.81,0.902,0.957}
\definecolor{daggreen}{RGB}{200, 214, 191}
\tikzstyle{founder}=[draw, ellipse, fill=yellow, inner sep=0pt,minimum size=1cm, double]
\tikzstyle{unknown} = [draw, ellipse, fill=yellow, inner sep=0pt,minimum size=1cm]
\tikzstyle{data} = [draw, fill=daggreen, inner sep=0pt,minimum size=1cm]
\tikzstyle{note} = [draw=none,fill=none,right]
\tikzstyle{state} = [draw, fill=stateblue, inner sep=0pt,minimum size=1cm]

\if@twoside
\else
\fi
\def \dsp{\def \baselinestretch{0.9}\large \normalsize}
\dsp

\newcommand{\tbeta}{\tilde{\beta}}

\newcommand{\Cov}{\mathrm{Cov}}

\providecommand{\nni}{\Delta}
\providecommand{\ksens}{\kappa_{\textrm{\scriptsize sens}}}
\providecommand{\kspec}{\kappa_{\textrm{\scriptsize spec}}}

\usepackage{enumitem}

\begin{document}

\def\spacingset#1{\renewcommand{\baselinestretch}%
{#1}\small\normalsize} \spacingset{1}


\if0\blind
{
  \title{A computationally efficient framework for realistic epidemic modelling through Gaussian Markov random fields}
  
\author[1,*,**]{Angelos Alexopoulos}
\author[3,2]{Paul Birrell}
\author[2,3]{Daniela De Angelis}

\affil[1]{\small Department of Economics, Athens University of Economics and Business, Greece}
\affil[2]{\small MRC Biostatistics Unit, University of Cambridge, UK}
\affil[3]{\small UK Health Security Agency, London, UK}

\affil[*]{\small The majority of the work was carried out at the MRC Biostatistics Unit, University of Cambridge}
\affil[**]{\small Corresponding author. E-mail: angelos@aueb.gr}
  \maketitle
} \fi

\if1\blind
{
  \bigskip
  \bigskip
  \bigskip
  \begin{center}
    {\LARGE\bf A computationally efficient framework for realistic epidemic modelling through Gaussian Markov random fields }
\end{center}
  \medskip
} \fi

\bigskip
\begin{abstract}
We tackle limitations of ordinary differential equation-driven Susceptible-Infections-Removed (SIR) models and their extensions that have recently be employed for epidemic nowcasting and forecasting. In particular, we deal with challenges related to the extension of SIR-type models to account for the so-called \textit{environmental stochasticity}, {\it i.e.} external factors, such as seasonal forcing, social cycles and vaccinations that can dramatically affect outbreaks of infectious diseases. Typically, in SIR-type models environmental stochasticity is modelled through stochastic processes. However, this stochastic extension of epidemic models leads  to models with large dimension that increases over time. Here we propose a Bayesian approach to build an efficient modelling and inferential framework for epidemic nowcasting and forecasting by using Gaussian Markov random fields to model the evolution of these stochastic processes over time and across population strata. Importantly, we also develop a bespoke and computationally efficient Markov chain Monte Carlo algorithm to estimate the large number of parameters and latent states of the proposed model. We test our approach on simulated data and we apply it to real data from the Covid-19 pandemic in the United Kingdom.
\end{abstract}

\noindent%
\vfill
\newpage
\spacingset{1.9} 

\section{Introduction}

A prompt health response to an epidemic relies on the ability to monitor its current state (nowcast) and forecast its evolution in real time. Traditionally, deterministic ordinary differential equation (ODE)-driven Susceptible-Infectious-Removed (SIR) type models \citep{bailey1971estimation,anderson1992infectious,jacquez1996compartmental} are used to approximate the transmission process of diverse infections (e.g. \cite{grenfell1992chance,keeling2002understanding}, \cite{lekone2006statistical}, \cite{birrell2011bayesian}, \cite{dukic2012tracking}).  In models of this type the population is subdivided into disease states with the flow between states described by an ODE system, which provides a mathematical expression for the \textit{intrinsic} transmission dynamics of the pathogen under consideration. However, infectious disease outbreaks are affected by \textit{external} stochastic factors, e.g., seasonal forcing, social cycles, vaccinations, or emergence of new variants, which cannot easily be incorporated in the ODE structure of SIR-type models. These additional epidemic drivers contribute to what is known as \textit{environmental stochasticity} (see, for example, \cite{dureau2013capturing} and \cite{ghosh2022differentiable} for related discussions). 
To account for such stochasticity, the standard SIR-type model can be extended by introducing a time-varying transmission rate parameter $\beta_t$ with a temporal evolution modelled through a stochastic process ({\it e.g.} Brownian motion). This approach has been adopted, among others, by \cite{dureau2013capturing}, \cite{cazelles2018accounting}, \cite{birrell2021real}, \cite{knock2021key} and \cite{bouranis2022bayesian}.  
Inference for these extended ODE systems is typically performed in a Bayesian framework, which is best suited to tackle the structural and practical identifiability of the model parameters \citep{chis2011structural} due to the lack of direct data on the infection process. 

The introduction of a stochastic process for $\beta_t$ poses a number of new challenges. 
Firstly, tracking the entire stochastic process results in a parameter space of high dimension that increases over time. To tackle the increase in dimensionality, it is common to assume that $\beta_t$ evolves according to a discrete-time  piecewise-constant stochastic process (see, {\it e.g.} \cite{cazelles2018accounting}, \cite{birrell2021real}, \cite{knock2021key} and \cite{bouranis2022bayesian}), introducing an approximation to the true continuous-time dynamics. Another typical assumption in models incorporating environmental stochasticity is that the stochastic processes that govern $\beta_t$ in different population strata ({\it e.g.} age and geographical location) are independent. This ignores any structure in the population under study and leads to unrealistic forecasts, resulting from not fully utilizing the available information. A further limitation of such models is related to the algorithm used in the Bayesian analysis. Often inference relies on adaptive MCMC algorithms \citep{flaxman2020estimating,birrell2024real}. Adaptive methods, however,  scale poorly and to  high dimensions \citep{rosenthal2011optimal}. On the other hand, more sophisticated MCMC algorithms that utilize the geometry of the target posterior (see {\it{e.g.}} \cite{girolami2011riemann}) carries a computational cost, may make their application impractical for practitioners interested in real time analysis.

In this paper, we tackle the above limitations, making two contributions. First, we build a generic epidemic modelling framework that accounts for environmental stochasticity, which includes previous models from \cite{dukic2012tracking} to \cite{birrell2021real} and \cite{bouranis2022bayesian} as special cases. The basic idea of our proposed modelling framework is to account for environmental stochasticity through Gaussian Markov Random Fields (GMRFs) \citep{rue2005gaussian}. We show how the use of a GMRF will provide greater control over the discretisation frequency, whilst permitting correlation and structure in transmission between population strata.
The second contribution is the development of an MCMC algorithm tailored for Bayesian inference for GMRFs.  
This is achieved by constructing a sampler that draws from the full conditional of the latent \textit{dynamic} states of the model given the \textit{static} parameters and from the full conditional of the parameters given the latent states. Although in typical applications the former is the most challenging step (see {\it e.g.} \cite{girolami2011riemann} and \cite{titsias2018auxiliary} for related discussions), in ODE-based epidemic models both steps are particularly problematic given the indirect and noisy nature of the data (see \cite{valderrama2019mcmc} for similar applications).


To sample from the full conditional of the static parameters we combine the adaptive Metropolis with global scaling sampler \citep{haario2001adaptive,andrieu2008tutorial} with a randomised blocking strategy \citep{chib2010tailored}. We first learn the covariance structure of the target distribution and then randomly allocate parameters to proposal blocks at each iteration. The adopted random blocking strategy aims to deal with situations where the linear correlation between the parameters is not the optimal criterion for block membership. To sample latent states we develop a proposal mechanism informed from their prior correlation structure. In particular, we build a Metropolis-Hastings (MH) step to update all the dynamic parameters simultaneously by constructing a proposal distribution based on the (informative) prior for the latent states. We demonstrate the better mixing properties of the proposed sampler compared to samplers used in a similar context through a simulation study.

We evaluate the proposed modelling approach by comparing its out-of-sample predictive performance against that of  pre-existing models, using both real data from the Covid-19 pandemic in the UK and simulated data. 

The paper is organised as follows.In Section \ref{sec:model} the stochastic extension of an SEIR (Susceptible-Exposed-infectious-Recovered) models is introduced. Section \ref{sec: Bayesian} presents the bespoke MCMC algorithm and in Section \ref{sec:case_study} we show results from the application of the proposed methods on real and simulated data. Finally, Section \ref{sec:conclusions} concludes with a discussion.



\section{A stochastic extension of the SEIR model}
\label{sec:model}


\subsection{The SEIR model}
\label{sec:SEIR}

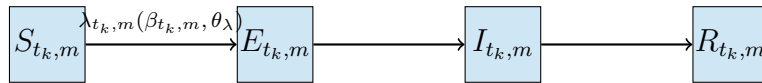
\begin{figure}[H]

      \begin{center}
          \begin{tikzpicture}[scale=0.9]
            \node[state] (St) {$S_{t_k,m}$};
            \node[state, right=of St] (Et) {$E_{t_k,m}$};
            \node[state, right=of Et] (It) {$I_{t_k,m}$};
            \node[state, right=of It] (Rt) {$R_{t_k,m}$};

            \draw[thick, ->] (St) -- (Et) node[midway,above,text=black] {\scriptsize $\lambda_{t_k,m}(\beta_{t_k,m},\theta_{\lambda})$};

            \draw[thick, ->] (Et) -- (It) node[midway,above] {$\phantom{\gamma}$};

            
            \draw[thick, ->] (It) -- (Rt) node[midway] {$\phantom{\rho}$};
          \end{tikzpicture}
      \end{center}
      \caption{Schematic diagram of simple SEIR model\label{fig:schematic_SEIR}}
    \end{figure}

We assume a closed population which, at time $t$ is stratified into $M$ \textit{strata}. Within stratum $m \in \{1,\ldots,M\}$, the population is partitioned into: a susceptible state $S_{t,m}$ which contains individuals who are at risk of becoming infected; an exposed state, $E_{t,m}$, including individuals who have been infected but are not yet infectious; an infectious state, $I_{t,m}$; and a removed/recovered state $R_{t,m}$. The total time spent in the $E$ and $I$ states are referred to as the latent and infectious  period respectively (see Figure \ref{fig:schematic_SEIR} for a schematic diagram of the model). We also assume that the system of ODEs that describes the SEIR model is evaluated at discrete times $t_k = k\delta$, $k=1,\ldots,K$, for some suitable time-step $\delta$. Under this framework the number of new infections in an interval $[t_{k-1},t_k)$ is
\begin{equation}\label{eq:nni}
    \nni_{t_k,m} = S_{t_{k-1},m}\lambda_{t_{k-1},m},
\end{equation} 
where $\lambda_{t_k,m} \equiv \lambda_{t_k,m}(\beta_{t,m},\theta_{\lambda})$ is the rate at which susceptible individuals become infected, parameterised through a time-dependent transmission parameter $\beta_{t_k,m}$ and a vector of \textit{static} transmission parameters $\theta_{\lambda}$ so that $\nni_{t_k,m} \equiv \nni_{t_k,m}(\beta_{t_k,m},\bm{\theta})$. 

\subsection{Environmental stochasticity by using GMRFs}
\label{sec:GMRFprior}

To account for external factors 
affecting disease dynamics, we assume that the time-dependent transmission parameter $\tbeta_{t_k,m} = \log(\beta_{t_k,m})$ evolves over time and across the different strata according to a GMRF on a lattice constructed by time points $t_k$ and strata $m$. Letting $\bm{\tbeta}_m = (\tbeta_{m,t_1},\ldots,\tbeta_{m,t_{\widetilde{K}}})^\top$, where $\widetilde{K}$ is the number of distinct values in $\bm{\tbeta}_m$, and $\bm{\tbeta} = (\bm{\tbeta}_1^\top,\ldots,\bm{\tbeta}_M^\top)^\top$, the Markov random field governing the joint distribution of $\bm{\tbeta}$ is determined by the full conditional distributions with densities $\pi(\tbeta_{m,t_k}|\bm{\tbeta}_{-m,-t_k})$ where $\bm{\tbeta}_{-m,-t_k}$ includes all the elements of the vector $\bm{\tbeta}$ except from $\tbeta_{m,t_k}$. Under the GMRF model a joint Gaussian prior distribution for $\bm{\tbeta}$ with some mean and precision (inverse covariance) matrix is defined. The precision matrix encodes the conditional correlation structure of the elements of $\bm{\tbeta}$; see for example in \cite{rue2005gaussian} for more details. To model the evolution of the log-transformed transmission parameters over time and across strata, we use the intrinsic GMRF (IGMRF) models in which the precision (inverse covariance) matrix of the joint (Gaussian) distribution of $\bm{\tbeta}$ is singular, {\it i.e}it does not have full rank (see Supplementary material for further details). Note also that this model specification can be also interpreted as a discretisation of correlated strata specific continuous time stochastic processes.

Typically the discretisation of the ODE system is achieved by setting $\delta$ to a small value. 
Notably, in contrast to the existing literature \citep{birrell2021real,bouranis2022bayesian}, in our proposed modelling approach we do not make any ex-ante assumptions to reduce the dimension of $\bm{\tbeta}$. We assume that $\tilde{K} = K/\delta_{\beta}$, where $\delta_{\beta}$ can be chosen according to the specific computational and/or modelling needs of the application.  Specifically, we assume that
$\bm{\tbeta}$ follows a Gaussian distribution with zero mean and precision (inverse covariance) matrix $\bm{Q}$ where 
\begin{equation}
\label{eq:Qmatrix}
\bm{Q} = \tau(\rho_{M}P_{M}\otimes I_{\widetilde{K}} + \rho_{time}I_{M}  \otimes P_{\widetilde{K}})
\end{equation}
where  $I_{\widetilde{K}}$ and $I_{M}$ are the $\widetilde{K} \times \widetilde{K}$ and $M \times M$ identity matrices respectively and $\otimes$ denotes the Kronecker product. The matrices $P_M$ and $P_{\widetilde{K}}$ encode the correlation structure across strata and time respectively, giving the precision matrix in \eqref{eq:Qmatrix} flexibility to model temporal and strata-specific dependencies. By choosing, for example, $P_{\widetilde{K}}$ to be the tridiagonal precision matrix of the random walk model and setting $P_M$ to be the identical matrix $I_{M}$ and  $\rho_M= \rho_{time}=1$,  we retrieve the structure assumed by \cite{birrell2021real}, \cite{knock2021key} and \cite{bouranis2022bayesian}. Independently of the choices for $P_{\widetilde{K}}$ and $P_M$, $\bm{\tbeta}$ is an IGMRF since $\bm{Q}$ is singular and the induced improper density of $\bm{\tbeta}$ can be written as  
\begin{equation}
    \label{eq:rw_dens}
    \pi(\bm{\tbeta}|\theta_{\bm{\tbeta}}) \propto \tau^{-\widetilde{K}M/2}\exp\left\{  -\frac{1}{2}\bm{\tbeta}^\top \bm{Q} \bm{\tbeta} \right\},
\end{equation}
where we condition on the vector with parameters $\theta_{\bm{\tbeta}}$ due to the dependence of the specified GMRF model on the parameters $\theta_{\bm{\tbeta}} = (\tau,\rho_M,\rho_{time})$.

\section{Bayesian inference}
\label{sec: Bayesian}

A big challenge in epidemic models is that inference typically can only rely on indirect data about new infections. This makes identifiability of parameters particularly challenging. In addition, data are generally collected from multiple, biased, sources  ({\it e.g.}  hospitalisations or deaths that only identify infections in individuals with greater frailty). A Bayesian framework is particularly suited to synthesise imperfect and indirect data; through appropriate specification of priors, allows identification of parameters that cannot be identified easily from indirect data; and enables easy translation of uncertainty in model parameters to epidemic quantities of key public and policy interest.

In this work, 
consistently with the data structures commonly used in epidemic studies \citep{flaxman2020estimating,bouranis2022bayesian,birrell2024real}, we assume the availability of surveillance data from multiple sources to estimate the parameters of interest in the SEIR-type models introduced in Section \ref{sec:model}. These data are typically linked to the transmission model by specifying appropriate observational models. In particular, let us assume that $y_{t_k,m}^j$, the observation at time $t_k$ for stratum $m$ from the $j$-th data source, follows a distribution with density $p(y_{t_k,m}^j|\mathcal{S}_{t_1:t_k,m},\bm{\tbeta}_{t_1:t_k,m},\bm{\theta})$, where $\mathcal{S}_{t_1:t_k,m}=(S_{t_1:t_k,m},E_{t_1:t_k,m},I_{t_1:t_k,m},R_{t_1:t_k,m})$, the subscript $t_1:t_k$ on a process implies that we condition on the values of the process at times $t_1,\ldots,t_k$ and $\bm{\theta} = (\theta_{\lambda},\theta_{\bm{\tbeta}})$ denotes all the parameters of the stochastically extended SEIR model. Then, if 
$$
\mathcal{D} = \big\{ y_{t_k, m}^j \mid k = 1, \dots, K; \, m = 1, \dots, M; \, j = 1, \dots, J \big\},
$$
with $J$ indicating the number of data sources, denotes all the available data,  the posterior of interest is  
\begin{equation}
    \label{eq:general_posterior}
    p(\bm{\theta},\bm{\tbeta}|\mathcal{D} ) \propto p(\mathcal{D}|\bm{\theta},\bm{\tbeta})\pi(\bm{\tbeta}|\bm{\theta})\pi(\bm{\theta}),
\end{equation}
where $\pi(\bm{\tbeta}|\bm{\theta})$ is specified through \eqref{eq:rw_dens} and $\pi(\bm{\theta})$ is the density of the prior distribution for the parameters $\bm{\theta}$ and $p(D|\bm{\theta},\bm{\tbeta})$ is specified through the densities $p(y_{t_k,m}^j|\mathcal{S}_{t_k,m},\beta_{t_k,m},\bm{\theta})$.

\subsection{Bayesian estimation through MCMC}
\label{sec:mcmc}

For inference we are interested in the posterior distribution in \eqref{eq:general_posterior}, which is typically very high-dimensional and intractable, requiring carefully-designed MCMC to generate posterior samples that can be used to estimate epidemic quantities of interest. In particular, we alternate sampling of the static parameters $\bm{\theta}$ and the dynamic parameters $\tbeta$ from their full conditional distributions $p(\bm{\theta}|\bm{\tbeta},\mathcal{D} )$ and $p(\bm{\tbeta}|\bm{\theta},\mathcal{D} )$ respectively. Both the sampling steps are challenging. 

\subsubsection{Sampling from $p(\bm{\theta}|\bm{\tbeta},\mathcal{D} )$}

To draw samples from $p(\bm{\theta}|\bm{\tbeta},\mathcal{D} )$ \cite{birrell2021real} use the the random walk MH (RW-MH) algorithm and Birrell et al (2024) the Adaptive Metropolis with global adaptive scaling (AMGS); see Algorithm 4 in \cite{andrieu2008tutorial}. In the latter, components of $\theta$ are blocked together based on convenience and updated according to an adaptive algorithm that progressively refined the proposal covariance matrix. Here, we extend this approach by integrating AMGS with the randomised blocking strategy proposed by \cite{chib2010tailored}. 

More precisely, we design a multiple-block MH algorithm,
where the parameters are grouped into several distinct blocks and each block of parameters is updated in sequence by a MH step, conditioned on the most current value of the parameters in the remaining blocks. 
It is common practice to form blocks consisting of highly correlated parameters, whereas parameters in different blocks are not strongly correlated. However, grouping the parameters based on the level of their correlation is not always the optimal criterion. This, for example, can happen when the shape of the correlation of the parameters changes over the MCMC iterations; see \cite{chib2010tailored} for a detailed discussion. Therefore, here we adopt the randomised blocking strategy proposed by \cite{chib2010tailored} to avoid a poor {\it a priori} choice of blocks. Our proposed MCMC algorithm to update the parameters in $\bm{\theta}$ is described in Algorithm \ref{alg:theta_sampling}. Notice that the algorithm requires as input the covariance matrix of $\bm{\theta}$; in our applications we estimate this in an adaptive manner during the burn-in period of our sampler; see Algorithm \ref{alg:mcmc} in the Appendix for more details.

\begin{algorithm}[H]
\caption{Randomized block sampler $p(\bm{\theta}|\bm{\tbeta},\mathcal{D} )$}
{\bf Input}: Covariance matrix $\Sigma = \Cov(\bm{\theta})$. 
\begin{algorithmic}[H!]
\State Divide randomly $\bm{\theta}$ in $P$ blocks $\bm{\theta}_1,\ldots,\bm{\theta}_P$. 
\For{$j=1$ {\bfseries to} $P$}
\State Sample $\bm{\theta}_j$ by using MH with proposal covariance matrix $\Sigma_j = \Cov(\bm{\theta}_j)$.
\EndFor
\end{algorithmic}
\label{alg:theta_sampling}
\end{algorithm}


\subsubsection{Sampling from $p(\bm{\tbeta}|\bm{\theta},D)$}
\label{sec:beta_sampling}

To draw samples from the posterior distribution with density 
$p(\bm{\tbeta}|\bm{\theta},\mathcal{D} )$ we need to design an MCMC algorithm that exploits the high autocorrelation
of the components of $\bm{\tbeta}$ and scales well as the dimension of $\bm{\tbeta}$ increases. We construct an MH step with a prior-informed proposal distribution to efficiently draw samples from the target $p(\bm{\tbeta}|\bm{\theta},\mathcal{D} )$; \cite{beskos2008mcmc} have shown that failing to be prior-informed leads to collapse of algorithms in high-dimensional problems due to acceptance rates becoming prohibitively low. Here we follow \cite{titsias2018auxiliary} and explore an idea that first appeared in the contribution made by Titsias (2011) to the discussion of \cite{girolami2011riemann}. 

To construct a prior-informed MH step 
that targets the density
\begin{equation}
\label{eq:target_beta}
        p(\bm{\tbeta}|\bm{\theta},\mathcal{D} ) \propto  \pi(\bm{\tbeta})\exp\{g(\bm{\tbeta};\bm{\theta},\mathcal{D} )\}, 
\end{equation}
where $\pi(\bm{\tbeta})$ is given by \eqref{eq:rw_dens} and $g(\bm{\tbeta},\bm{\theta};\mathcal{D} )$ denotes the log-likelihood function, 
we follow the auxiliary variables approach \citep{titsias2018auxiliary}. We augment the target distribution with the auxiliary variable $\bm{u} \sim N(\bm{\tbeta},(c^2/2)I_{\delta_{\beta}})$, i.e., $\bm{u}$ is equal to 
$\bm{\tbeta}$ with some added Gaussian noise. Then, we draw samples from the distribution with density 
\begin{align}
\label{eq:aug_target}
p(\bm{\tbeta},\bm{u}|\bm{\theta},\mathcal{D}) &= p(\bm{u}|\bm{\tbeta})p(\bm{\tbeta}|\bm{\theta},\mathcal{D}) \\
&\propto N(\bm{u}|\tbeta,(c^2/2)I_{\delta_{\beta}})\exp\bigg\{ -\frac{1}{2}\bm{\tbeta}^\top Q \bm{\tbeta} \bigg\}\exp\{g(\bm{\tbeta};\bm{\theta},\mathcal{D})\}, 
\end{align}
by alternating the following two steps:
\begin{itemize}
    \item[(i)] Sample from  $p(\bm{u}|\bm{\tbeta},\bm{\theta},\mathcal{D} ) \equiv p(\bm{u}\lvert\bm{\tbeta})$.
    \item[(ii)] Sample from $p(\bm{\tbeta}|\bm{u},\bm{\theta},\mathcal{D} )$.
\end{itemize}
From \eqref{eq:aug_target} it is clear that 
$p(\bm{u}|\bm{\tbeta},\bm{\theta},\mathcal{D} ) = N(\bm{u}|\tbeta,(c^2/2)I_{d_{\beta}})$ whereas 
$$
p(\bm{\tbeta}|\bm{u},\bm{\theta},\mathcal{D} ) \propto N\bigg(\bm{\tbeta}|\frac{2}{c^2}A^{-1}\bm{u},A^{-1}\bigg)\exp\{g(\bm{\tbeta};\bm{\theta},\mathcal{D} )\},
$$  
where $\bm{A} = (Q + \frac{2}{c^2}I_{d_{\beta}})$, is not of standard form.
Therefore, we need a RW-MH step to conduct (ii). However, \eqref{eq:aug_target} suggests that a 
proposal for $p(\bm{\tbeta}|\bm{u},\bm{\theta},\mathcal{D} )$ is the prior distribution $p(\tbeta \lvert \bm{u}) = N\bigg(\bm{\tbeta}|\frac{2}{c^2}\bm{A}^{-1}u,A^{-1}\bigg)$. The acceptance ratio of the developed RW-MH step will only rely on the likelihood assumed for the observed data evaluated at the proposed and current value of $\bm{\tbeta}$, i.e,
a proposed value $\bm{\tbeta}^\star$ will be accepted with probability 
\begin{equation}
    \label{eq:acc_ratio_beta}
    \min\big(1,\exp\{g(\bm{\tbeta}^\star;\bm{\theta},\mathcal{D} )-g(\bm{\tbeta};\bm{\theta},\mathcal{D} )\}\big).
\end{equation}
This implies that the proposed RW-MH step is invariant with respect to $\pi(\bm{\tbeta})$ which is a desired property of MCMC algorithms \citep{neal1998regression,beskos2008mcmc}. Moreover, we note that although $\bm{Q}$ is singular the matrix $\bm{A}$ is invertible which is also an essential property of the proposed RW-MH algorithm. Algorithm \ref{alg:tp} summarizes the steps of the proposed scheme for sampling from $p(\bm{\tbeta},\bm{u}|\bm{\theta},\mathcal{D} )$; notice that our final aim of drawing samples from $p(\bm{\tbeta}|\bm{\theta},\mathcal{D} )$ is achieved since this is a marginal the joint target defined in \eqref{eq:aug_target}. The developed RW-MH step is  a special case of the gradient based sampler developed by \cite{titsias2018auxiliary} when the gradient term of their proposal mechanism is turned off to limit the computational cost of calculating gradients from ODE-based models.

\begin{algorithm}[H]
\caption{Auxiliary sampler to draw samples from $p(\bm{\tbeta},\bm{u}|\bm{\theta},\mathcal{D} )$}
\begin{algorithmic}[H!]
\State $1$. Sample $\bm{u} \sim N_{\delta_{\beta}}(\bm{\tbeta},(c^2/2)I_{\delta_{\beta}})$.
\State $2$. Propose $\bm{\tbeta}^\star \sim  N\bigg(\frac{2}{c^2}A^{-1}\bm{u},A^{-1}\bigg)$ and accept them according to the Metropolis-Hastings ratio 
in \eqref{eq:acc_ratio_beta}
\end{algorithmic}
\label{alg:tp}
\end{algorithm}

\section{A case study: Covid-19 in the UK}
\label{sec:case_study}

To illustrate the proposed epidemic model as well as the new computational methods for Bayesian inference we take as an example the recent Covid-19 pandemic in the UK. We conduct a simulation study to compare the proposed modelling and computational framework with existing approaches; and  apply the proposed methodology to real data to evaluate the predictive ability of the extended model.

We start from the stratified SEIR-type model specified in Section \ref{sec:model}. Following \cite{birrell2021real,birrell2024real} we assume that the population is stratified into $M=7$ geographical strata defined by National Healthcare System (NHS) regions; and $n_A=8$ age groups ($<1$, 1--4, 5--14, 15--24, 25-44, 45--64, 65--74, $\geq$75). Within each region, the infection dynamics are governed by a system of discretised ODEs where the population of individuals is partitioned for each age group into the disease states $S_{m,t_k, i}$, $E_{m,t_k, i}$ and $I_{m,t_k, i}$ , $k=1,\ldots,K$, $m=1,\ldots,M$, for $M$ regions, and $i=1,\ldots,n_A$ the age group; the discretisation interval for the ODE system is controlled by the parameter $\delta$ defined in Section \ref{sec:model} and here we set $\delta= 1/2$ to be sufficiently small relative to the latent and infectious periods. The average period spent in the exposed and infectious states are given by the parameters $d_L$ and $d_I$ respectively; and $\lambda_{m,t_k,i}$ is the time- and age-varying rate at which susceptible individuals become infected; see also Figure \ref{fig:schematic_SEIR} for the schematic representation of the SEIR model.  The rate $\lambda_{m,t_k,i}$ depends the probability of a susceptible individual in region $m$ of age group $i$ being infected by an infectious individual in age group $j$ at time $t_k$ denoted by $b^{t_k}_{m,ij}$ and this probability is assumed to be driven by the unobserved process $\tbeta_{m,t_k}$ as well as from the expected number of contacts between individuals in age groups $i$ and $j$ within a single time unit $t_k$ denoted by $C^{t_k}_{m,ij}$, through the function
\begin{equation}\label{eqn:bij}
  b^{t_k}_{m,ij} = h(\tbeta_{m,t_k},C^{t_k}_{m,ij};\bm{\theta}),
 \end{equation} 
which expresses the dependence of $b^{t_k}_{m,ij}$ on static parameters $\bm{\theta}$, some of which appear in the system of ODEs that governs the dynamics of infection in the SEIR model (see Section \ref{sec:data} for the elements of $\bm{\theta}$ as well as the Appendix for more details about modelling the infectious probabilities in a SEIR model). Note that $ b^{t_k}_{m,ij}$ is of great epidemiological interest since it is deterministically related to the evolution of the reproduction number (see the Appendix for details). The parameters $\tbeta_{m,t_k}$ evolve over time and across the different regions according to the GMRF model introduced in Section \ref{sec:GMRFprior}.

\subsection{Data and likelihood}
\label{sec:data}

The set of available data are 
\begin{equation*}
    \mathcal{D} = \left\{\left(y^d_{m,t_k,i}, y^s_{m,t_k,i}\right)\right\}_{k=1}^K
\end{equation*}
and consist of age- and region- specific counts of deaths, $y^d_{m,t_k,i}$; and serological data indicating the number of blood sera samples testing positive for the presence of antibodies indicative of past infection, $y^s_{m,t_k,i}$ out of $n_{m,t_k,i}$ samples tested. 
Following \cite{birrell2021real} we assume $y^d_{m,t_k,i}$ on day $t_k$ follows a negative binomial distribution
\begin{equation}
\label{eq:negbinom_lik}
y^d_{m,t_k,i}|\bm{\theta},f \sim \mathrm{NegBin}\left(\mu_{m,t_k,i}, \eta\right),
\end{equation}
where $ \mu_{m,t_k,i} = p_i \sum_{l=0}^k f_{k - l} \nni_{m,t_l, i}$, $\nni_{m,t_l, i}$ as in \eqref{eq:nni}, $f$ denotes an assumed-known cumulative distribution function of the time from infection to death from the disease under study (here Covid-19), $p_i$ is an age-specific infection-fatality ratio and $\bm{\theta}$ denotes all the non-dynamic parameters of the model. Here $\eta$ is a dispersion parameter such that the expected value and the variance of the negative binomial distribution in \eqref{eq:negbinom_lik} are $\mu_{m,t_k,i}$ and 
$\mu_{m,t_k,i}\left(1 + \eta\right)$ respectively. In addition, if on day $t_k$, $n_{m,t_k,i}$ blood samples are taken from individuals in region $m$ and age-group $i$, and the observed number of positive tests is $y^s_{m,t_k,i}$, then
\begin{equation}
\label{eq:binom_lik}
y^s_{m,t_k,i}| \bm{\theta}   \sim \mathrm{Bin}\left(n_{m,t_k,i}, \ksens \left(1 - \frac{S_{m,t_k,i}}{N_{m,i}}\right) + \left(1 - \kspec\right)\frac{S_{m,t_k,i}}{N_{m,i}}\right),
\end{equation}
where $\ksens$ and $\kspec$ are the sensitivity and the specificity of the serological testing process respectively,  $N_{m,i}$ is the total population in age-group $i$ and region $m$ whereas 
$$
\bm{\theta} = (\eta,d_I,\ksens,\kspec,\{p_i\}_{i=1}^{n_A},\{z_{m,1},z_{m,2},z_{m,3},\psi_m,\ell_{0,m}\}_{m=1}^M),
$$ 
$z_{m,1}, z_{m,2}$ and $z_{m,3}$ are multipliers of the number of contacts $C^{t_k}_{m,ij}$ and $\psi_m,\ell_{0,m}$ are initial parameters of the ODE system that underlies the SEIR model (see the Appendix for more details including  the prior distributions used for the elements in $\bm{\theta}$).

\subsection{Simulation study}
\label{sec:sim_study}

To evaluate the performance of the proposed methodology we perform an extensive simulation study. 
As already noted, both the system of ODEs, describing the population's flows between the different states of the SEIR model, and the stochastic process used to model the environmental stochasticity are continuous time processes that have been discretised to facilitate inference, with the discretisation interval controlled by the parameter $\delta$ in the case of the ODE and $\delta_{\beta}$ for the stochastic process $\bm{\tbeta}$. In \cite{birrell2021real}  $\delta =1/2$ was reasonable for modelling the evolution of the Covid-19 pandemic in the UK. However, setting also $\delta_{\beta}=1/2$ would hugely increase the dimension of the parameter space of the resulting model. In fact, by assuming that the estimation of the model is performed using daily data over $K$ days, then the dimension of the dynamic parameters is $KM/\delta_{\beta} = 2KM$. 
To deal with the curse of the dimensionality it is common to choose larger $\delta_{\beta}$, dramatically decreasing the number of parameters to be estimated. 

We consider two simulation scenarios using datasets of the type defined in Section \ref{sec:data}, 
using $K = 100$ days, $M=7$ geographical regions and $n_A =8$ age groups. Both scenarios generate $\bm{\tilde{\beta}}$ from the GMRF model in Section \ref{sec:GMRFprior} by setting $\delta = \delta_{\beta}= 1/2$. The simulated datasets differ only in the specification of the precision matrix $\bm{Q}$ in \eqref{eq:Qmatrix}. In scenario A we assume that the matrix $P_K$ and $P_M$ in \eqref{eq:Qmatrix} are both the tridiagonal matrix of the random walk model; see in the Appendix for their specific form. This implies that the regions are ordered and, at any time, conditional on the $\bm{\tilde{\beta}}$ for other regions, the region specific process $\bm{\tilde{\beta}}_m$ will, a priori, behave as a random walk.
In data generation mechanism B we relax the assumption of dependence between regions and we assume that $P_M = \mathcal{I}_M$, the $M \times M$ identity matrix. Finally, for both scenarios A and B we set $\rho_M =0.5$ and $\rho_{time} =1.5$. To assess the predictive performance of the proposed modelling approach, we split the simulated datasets into training and test sets, comprising the first $86$ days and the remaining $14$ days, respectively. See the Appendix for the values of the parameters $\bm{\theta}$ and $\bm{\beta}$ used in the simulation. In the following Sections we present a comparison between three models based on their goodness-of-fit and their predictive ability. 


\subsubsection{Models under comparison}

Table \ref{tab:data_mechanism_and_models} displays the two data generation mechanisms and the three models (a)--(c)  to be compared in the simulation study. 
All three models represent a mis-specification of the data generation process, although (a) differs from A only in the assumed value of $\delta_{\beta}$, 
assuming daily, rather than half-daily change points for $\bm{\tbeta}$. Model (b) and data generation mechanism B differ similarly. Finally, to compare our modelling framework with a model that is a standard choice to account for environmental stochasticity ({\it e.g.}see \cite{flaxman2020estimating}, \cite{birrell2021real} and \cite{bouranis2022bayesian}), we also consider model (c) for which $\delta_{\beta} =14$, {\it i.e}, the stochastic process that underpins the parameters in $\bm{\tbeta}$ is kept constant over a 14 day time interval in each region.

\begin{table}[H]
    \centering
    \caption{Specification of the stochastic processes governing the parameters in $\bm{\tbeta}$ for the models used as data generation mechanisms as well as for the models fitted to the simulated datasets.}
    \begin{tabular}{cccccc}
        \toprule
        \textbf{Data Generation Mechanism} & \textbf{$\delta_{\beta}$} & \textbf{$\delta$} & \textbf{$P_M$} & \textbf{$P_K$} \\
        \midrule
        A & 1/2 & 1/2 & Tridiagonal & Tridiagonal \\
        \midrule
        B & 1/2 & 1/2 & $I_M$ & Tridiagonal \\
        \bottomrule
        \textbf{Model fitted to simulated data} &  &  &  &  \\
        \midrule
        (a) & 1 & 1/2 & Tridiagonal & Tridiagonal \\
        \midrule
        (b) & 1 & 1/2 & $I_M$ & Tridiagonal \\
        \midrule
        (c) & 14 & 1/2 & $I_M$ & Tridiagonal \\
        \bottomrule
    \end{tabular}
    \label{tab:data_mechanism_and_models}
\end{table}

The model comparisons made below are all based on averages calculated across ten replicates of scenarios A and B, with the data generated under each denoted $D^{\textrm{sim}}_u$, $u = 1, \ldots, 10$.

\subsubsection{Goodness-of-fit}

\begin{figure}[H]
    \centering
    \includegraphics[scale=0.25]{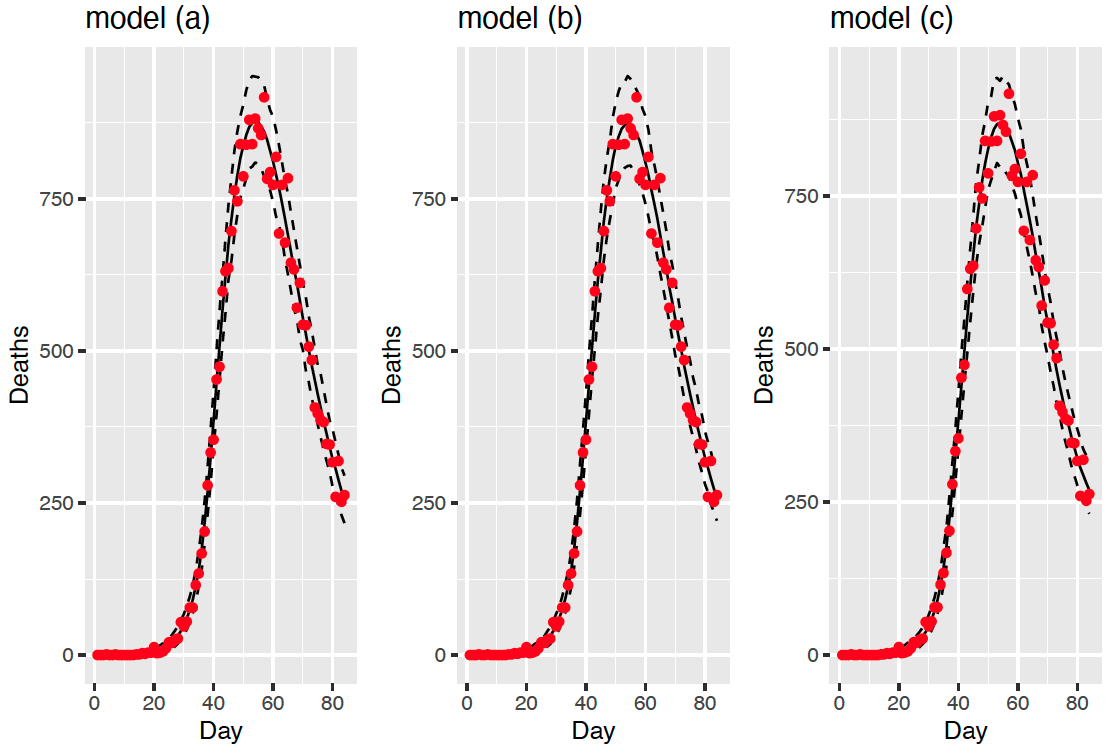}
    \caption{Posterior means (Black dashed lines) and $95\%$ credible intervals of predictive distributions for the total number of deaths for the data generation mechanism A. The red dots represent the simulated number of daily deaths.}
    \label{fig:gof_sim}
\end{figure}

Inference is carried out by sampling from the target posteriors $p(\bm{\theta},\bm{\tbeta}|D^{\textrm{sim}}_u)$ using the novel MCMC algorithm of Algorithm \ref{alg:mcmc}, implemented in bespoke R and C++ code by using the \texttt{Rcpp} package. We run the MCMC for 300,000 iterations and we discard the first 100,000 as a burn-in. The computational cost was different for different models with $1000$ iterations requiring 316, 303 and 68 seconds for model (a), (b) and (c), respectively. Figure \ref{fig:gof_sim} presents the simulated total number of daily deaths over the $K=86$ days from the first simulated dataset ($u=1$) under the data generation mechanism A, together with  the $95\%$ credible intervals of the corresponding posterior predictive distributions. Similar goodness-of-fit is also found when applying the models to data generated using mechanism B (see Appendix). 

\subsubsection{Predictive performance}

\begin{figure}[H]
    \centering
    \includegraphics[scale=0.65]{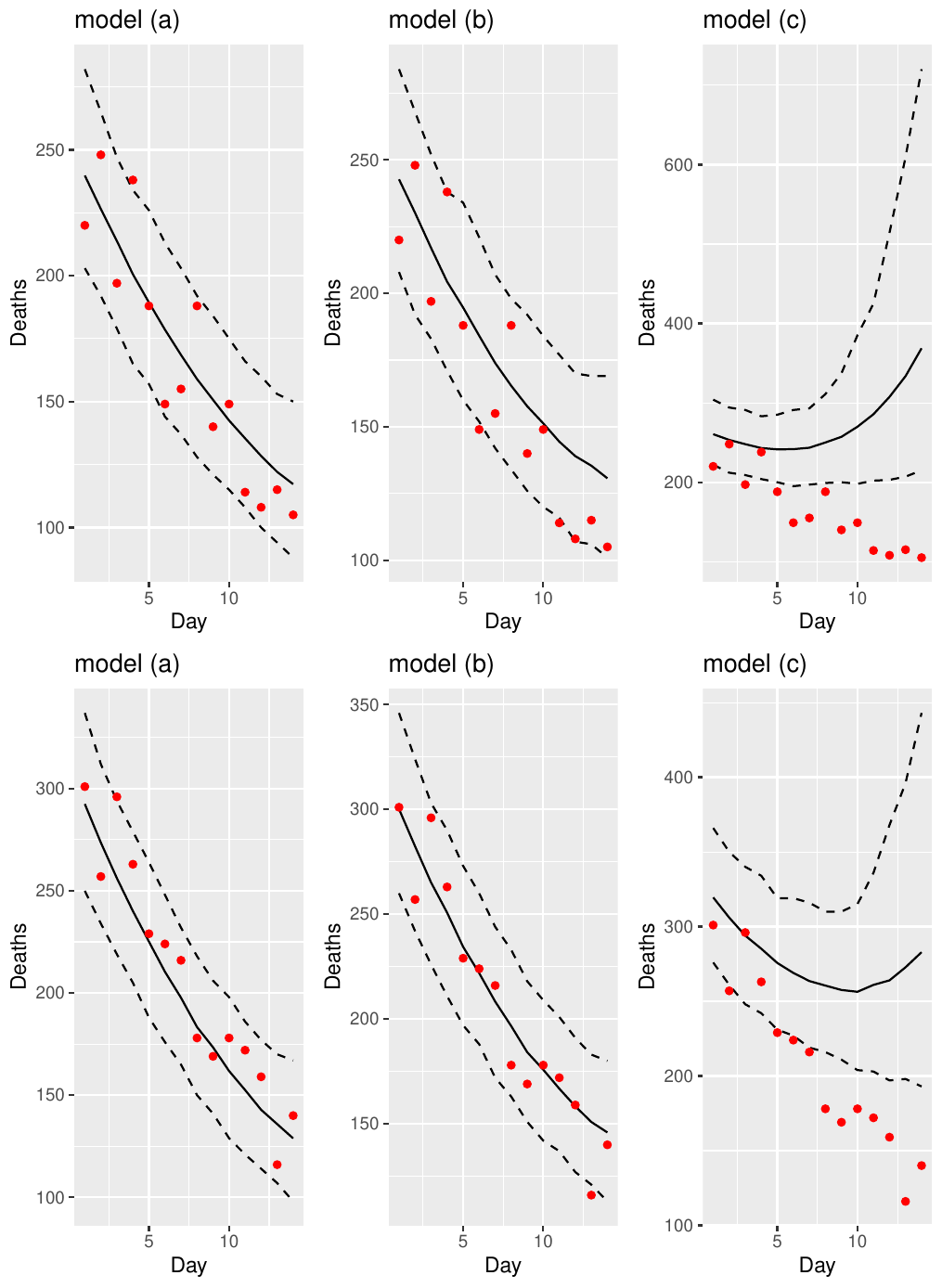}
    \caption{Posterior means (black dashed lines) and $95\%$ credible intervals of the posterior predictive distribution for the total number of deaths in each of the $14$ out-of-sample days. The red dots represent the true number of daily deaths. Top: simulated scenario A with correlated random walk processes across regions. Bottom: simulated scenario B with independent random walk.}
    \label{fig:preds_sim}
\end{figure}

For each model, we examine its out-of-sample predictive ability by drawing the number of deaths $\{y^d_{m,t_{K+\ell},i}\}$ for $\ell=1,\ldots,14$ from their posterior predictive distribution with density 
\begin{equation}
    \label{eq:post_pred_beta}
    p(\bm{\tbeta}_{t_{(K+1):(K+\ell)}}|D) = \int p(\bm{\tbeta}_{t_{(K+1):(K+\ell)}}|\bm{\tbeta}_{t_{1:K}},\bm{\theta})p(\bm{\tbeta}_{t_{1:K}}|\bm{\theta},D)d\bm{\tbeta}_{t_{1:K}},
\end{equation}
where $\bm{\tbeta}_{t_\ell} = (\tbeta_{1,t_\ell},\ldots,\tbeta_{M,t_\ell})^\top$ denotes the state of the random walk process at the $\ell$th out-of-sample time point and $\bm{\tbeta}_{1:t_\ell}=(\bm{\tbeta}_{1},\bm{\tbeta}_{2},\ldots,\bm{\tbeta}_{t_\ell})$. 
Figure \ref{fig:preds_sim} displays the means and the $95\%$ credible intervals of the posterior predictive distribution for the total number of daily deaths in the test dataset for both data generation mechanisms A and B and for $u=1$. It is clear from visual inspection of the figure that models (a) and (b) produce accurate predictions for the out-of-sample observations whereas the predictions from model (c) lose accuracy after the first few time points. 
Additionally, in scenario A where the random walk processes are simulated to be correlated across geographical regions, model (b), which assumes independence, produces wider prediction intervals.

To quantify the differences in the predictions obtained from the competing models we calculated interval scores and root mean squared errors (RMSE) for each model, both of which are minimised by a good probabilistic forecast. All the Figure and Tables presented in the rest of the section display averages across the $10$ simulated datasets for each of the data generation mechanisms A and B. 
We employed the R package \texttt{scoringRules} \citep{jordan2017evaluating} to use posterior samples to calculate the interval scores. 

\begin{figure}[H]
     \centering
     \begin{subfigure}[b]{0.45\textwidth}
         \centering
         \includegraphics[scale=0.3]{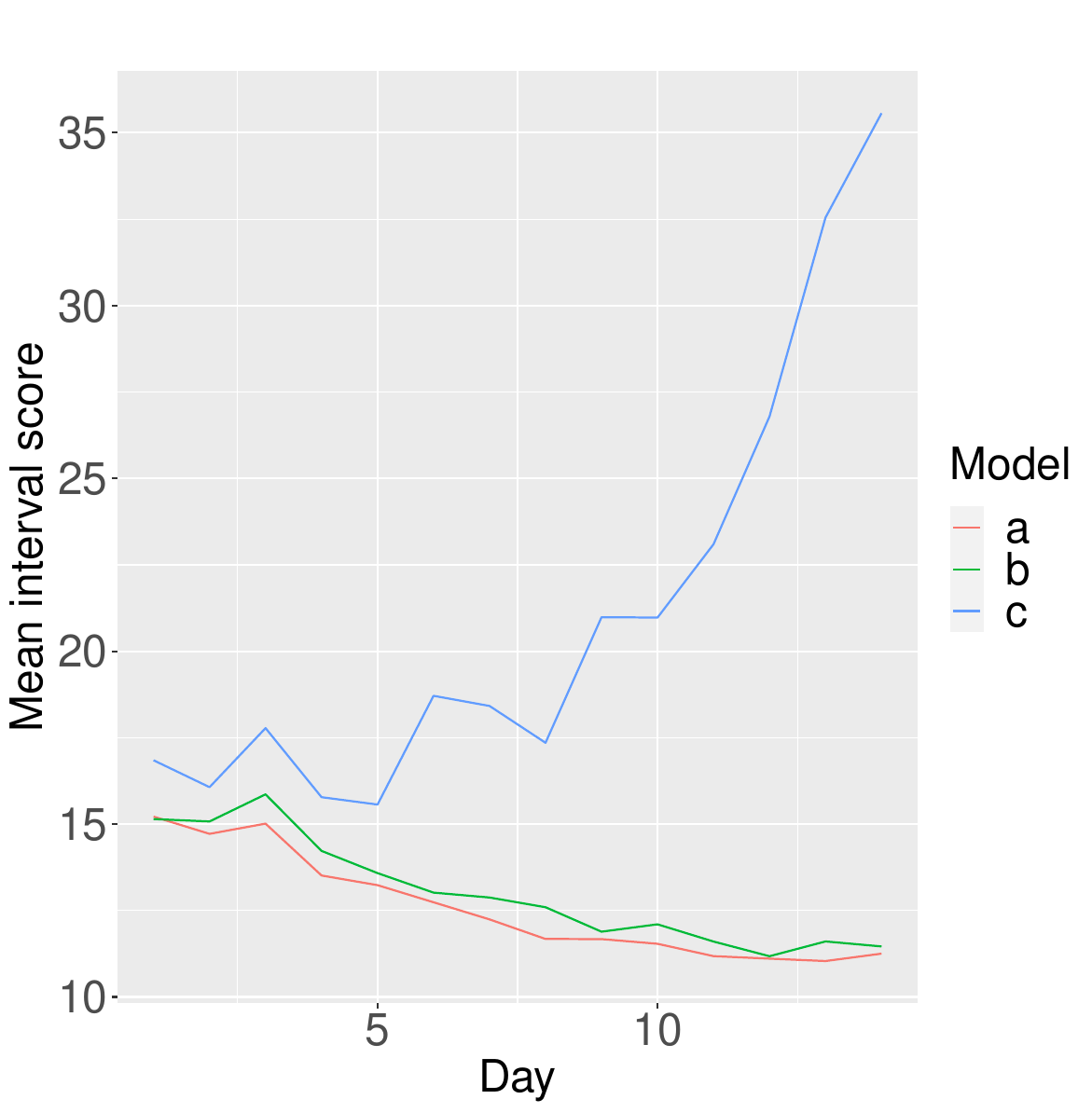}
         \label{fig:depend}
     \end{subfigure}
     \hfill
     \begin{subfigure}[b]{0.45\textwidth}
         \centering
         \includegraphics[scale=0.3]{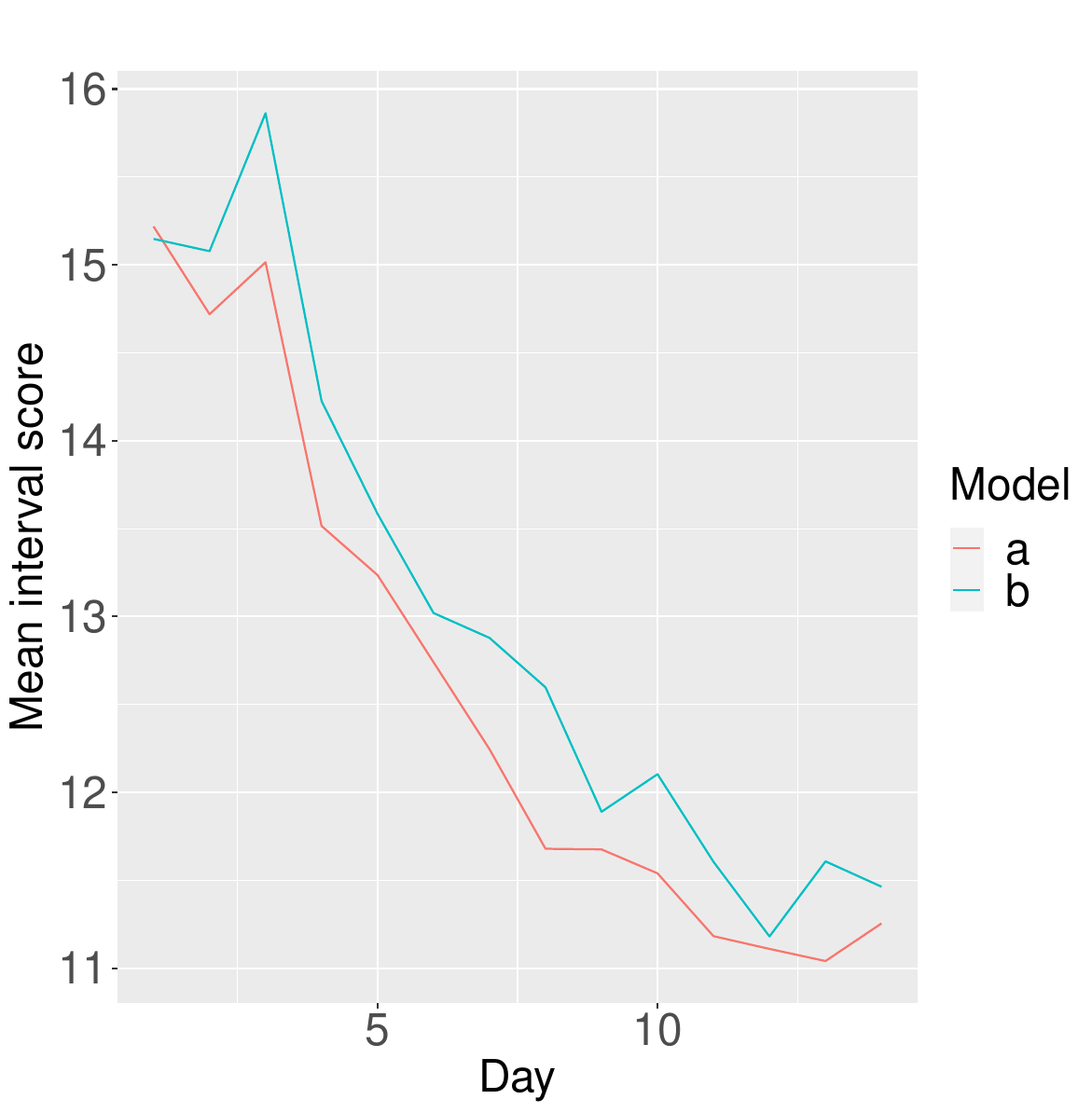}
         \label{fig:independ}
     \end{subfigure}
    \caption{Mean interval scores across seven regions for scenario A where the simulated random walk processes are independent across the regions. Left: comparison for the models (a) with daily changes in the states of the random walk process $\tbeta$ and the region-specific random walk process are correlated, (b) with daily changes $\tbeta$ and the region-specific random walk process are independent and (c) with bi-weekly changes in the states of $\tbeta$ and independent random walk processes across regions. Right: interval scores only for models (a) and (b).}
        \label{fig:interval_scores_A}
\end{figure}

Figure \ref{fig:interval_scores_A} presents mean (across the 7 regions) interval scores of the posterior predictive distribution of $y^d_{m,t_{K+\ell}} = \sum_{i=1}^8 y^d_{m,t_{K+\ell},i}$ calculated by fitting models (a), (b) and (c) to data simulated under data generation mechanism A, where the region-specific random walk processes are assumed correlated and $\delta_{\beta}=1$. It is clear, again, that model (c) has dramatically worse performance than models (a) and (b), due to the much more accurate discretisation scheme ($\delta_{\beta}=1$) of the random walk process $\bm{\tilde{\beta}}$ than in model (c) where the choice of $\delta_{\beta}$ corresponds to bi-weekly piecewise constant paths of the random walk process. By comparing the mean interval scores of models (a) and (b) we conclude that model (a) has slightly better predictive performance than model (b); this was expected since model (a) in this case satisfies the assumptions of the data-generating mechanism. Similar conclusions are made from the visual inspections of interval scores in cases with different levels of correlation between the regional random walk processes; see in the Appendix for more details where we present the mean interval scores corresponding to data generation mechanism B where we have simulated independent random walk processes in each region. Similarly to the case of data generation mechanism A, model (c) has the worst performance in terms of mean interval scores. 
Although model (a) assumes correlation between neighbouring regions, its prediction intervals improve slightly on the scores corresponding to model (b), the data generating model. Finally, Tables \ref{tab:sim_tab2} and \ref{tab:sim_tab1} (see in the Appendix for the latter), which display region-specific RMSEs for the three models under comparison for data generation mechanisms B and A respectively, indicate that model (c) provides less accurate point estimates of daily deaths for all the regions whereas models (a) and (b) have similar RMSEs in both scenarios.

Importantly, the results of our simulation studies provide evidence that the error in the discretisation of continuous time stochastic transmission processes commonly used in the literature has an important effect in prediction of key epidemic quantities. 
Finally, we evaluate the performance of the MCMC algorithm of Section \ref{sec:mcmc}, comparing it with MCMC methods that have been recently used to draw samples from epidemic models that are special cases of the proposed modelling framework. Tables \ref{tab:essC} and \ref{tab:essA} in the Appendix provide evidence that the developed MCMC sampler outperforms in terms of sampling efficiency the adaptive MCMC algorithm \citep{andrieu2008tutorial} which is routinely used for Bayesian inference in epidemics \citep{flaxman2020estimating,birrell2021real,bouranis2022bayesian}. Interestingly, the constructed sampler has better sampling properties for all the models compared to the previous Sections. Interestingly, although the differences are quite small in the case of model (c) where the parameter space is kept to low dimensions, in the case of models (a) and (b) the MCMC methods that are currently utilised to draw samples from target distributions in epidemic models (see e.g. Birrell et al., 2024) are struggling to explore the parameter space since their acceptance rate is prohibitively low; see the Appendix for more details. Therefore, the developed computational framework which allows for accurate discretisation schemes offers the opportunity to fit epidemic models that provide much more accurate projections for epidemic quantities of interest.

\begin{table}[H]
\caption{Region specific RMSEs for the models (a), (b) and (c) in the case of simulated scenario A where the random walk processes are correlated for different regions.}
\centering
\begin{tabular}{rrrr}
  \hline
 & (a) & (b) & (c) \\ 
  \hline
Region 1 & 50.36 & 50.75 & 52.73 \\ 
Region 2 & 81.80 & 83.74 & 359.99 \\ 
Region 3 & 51.99 & 53.59 & 54.09 \\ 
Region 4 & 182.34 & 188.13 & 272.32 \\ 
Region 5 & 18.08 & 18.17 & 64.68 \\ 
Region 6 & 52.93 & 52.76 & 54.19 \\ 
Region 7 & 131.19 & 134.39 & 185.03 \\ 
   \hline
\end{tabular}
\label{tab:sim_tab2}
\end{table}


\subsection{Real data application}
\label{sec:real_data}

We modelled the dynamics of the Covid-19 epidemic in the UK by relying on the following three types of data:  UKHSA's line-listing of deaths reported to have occurred within 60 days of a lab-confirmed PCR test; serological data giving the proportion of blood samples submitted to the National Health Service Blood and Transplant  service that tested positive for the presence of antibodies; and mobility indices derived from Google mobility, the UK Time-Use survey and data on school attendances from the UK Department for Education \citep{vLeS20}. All the data were stratified by $M = 7$ regions defined the National Health System (NHS) and $n_A = 8$ age groups ($<1$, 1--4, 5--14, 15--24, 25--44, 45--64, 65--74, $\geq75$). The study ranges from 2020-02-16 to 2020-06-16 ($121$ days).


\subsubsection{Bayesian estimation}

\begin{figure}[t]
    \centering
    \includegraphics[scale=0.3]{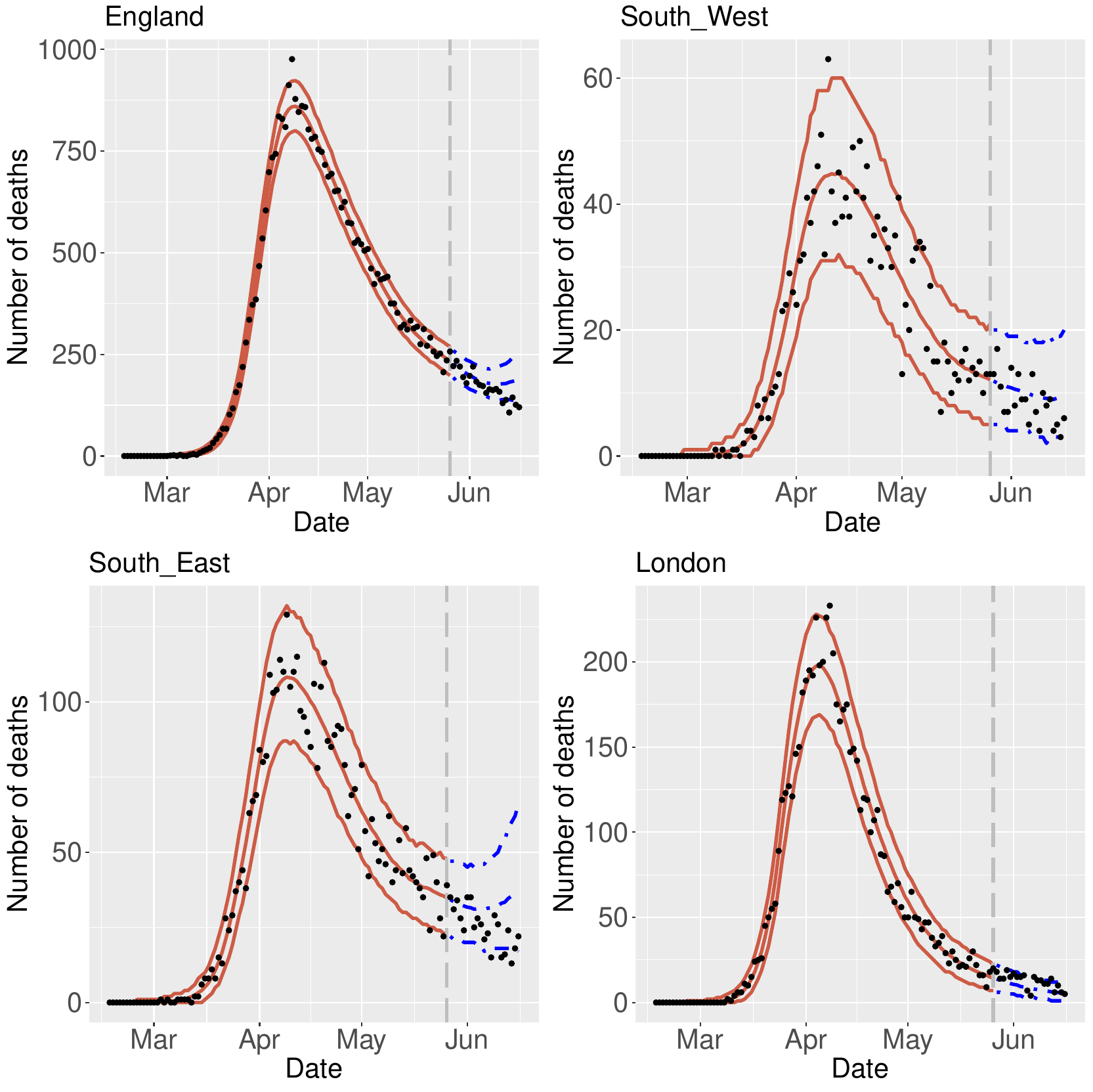}
    \caption{Red solid lines: $95\%$ credible intervals and means of the posterior distribution for the number of deaths estimated by using data from 2020-02-16 to 2020-05-27 for the whole England as well as for the regions South West, South East and London. Blue dotted lines: $95\%$ credible intervals and means for the posterior predictive distribution of the number of deaths from 2020-02-17 to 2020-06-16. Black dots: true number of deaths. The gray dashed line indicates the last day of the data used for the estimation of the model.}
    \label{fig:gof1}
\end{figure}

\begin{figure}[H]
    \centering
    \includegraphics[scale=0.3]{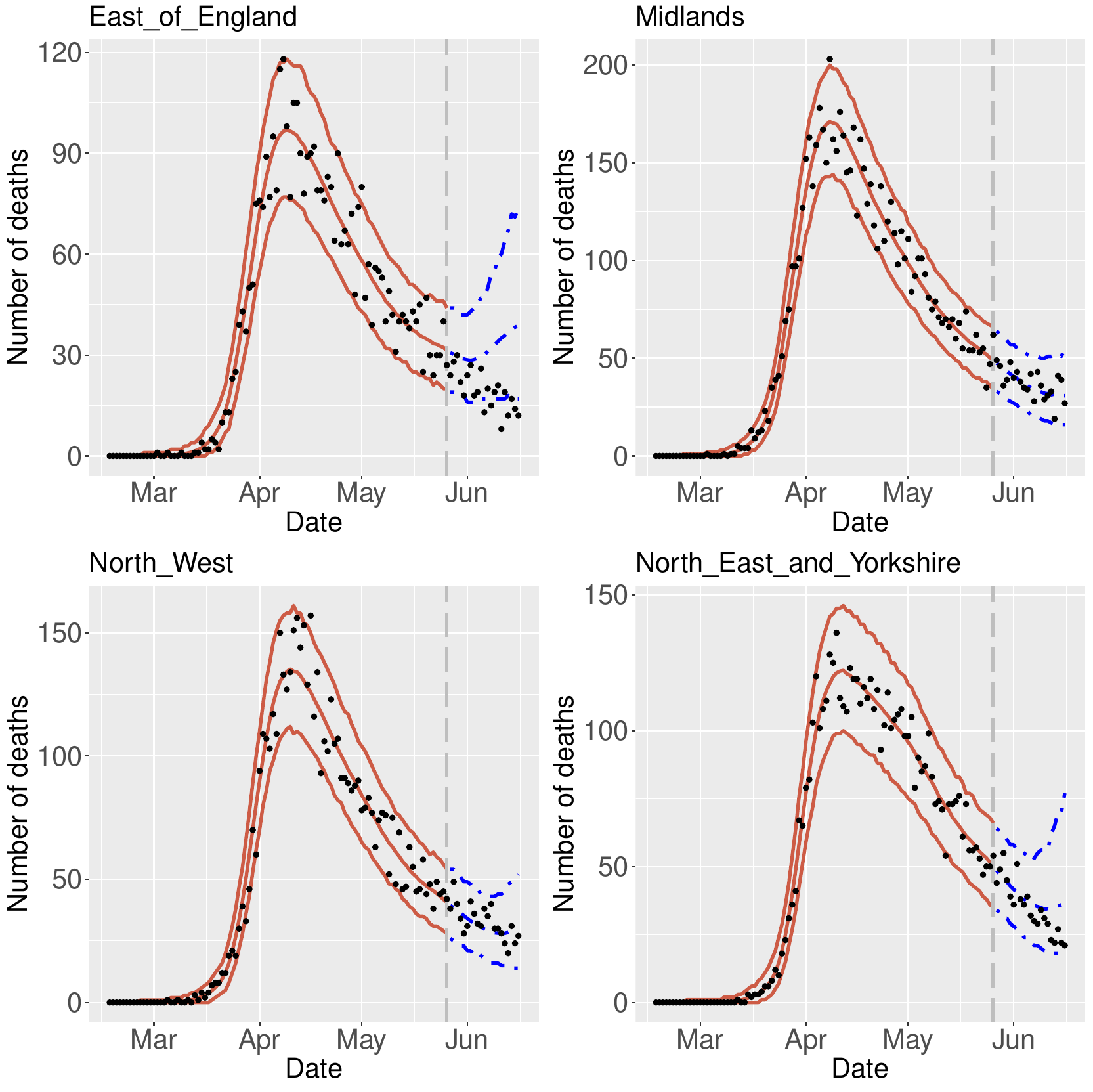}
        \caption{Red solid lines: $95\%$ credible intervals and means of the posterior distribution for the number of deaths estimated by using data from 2020-02-16 to 2020-05-27 for the regions East of England, Midlands, North West and North East and Yorkshire. Blue dotted lines: $95\%$ credible intervals and means for the posterior predictive distribution of the number of deaths from 2020-02-17 to 2020-06-16. Black dots: true number of deaths. The gray dashed line indicates the last day of the data used for the estimation of the model.}
    \label{fig:gof2}
\end{figure}

To test how the proposed modelling and computational framework behaves when challenged with real data we used the observations up to $K=100$ days in the described dataset to obtain posteriors for both parameters and latent states and then used the remaining $21$ days to evaluate forecasts. Figures \ref{fig:gof1} and \ref{fig:gof2} depict the $95\%$ credible intervals and the means of the posterior predictive distributions for the full time series of deaths in each region. Firstly, the Figures show an efficient reconstruction of the observed epidemic curve, with only a few of the $100$ in-sample observations falling outside of posterior credible intervals. Secondly, the model provides accurate predictions for the number of future deaths since the vast majority of the $21$ out-of-sample observations in each one of the NHS regions as well as in the whole of England are within the posterior predictive intervals.

\subsubsection{Forecasts evaluation}
\label{sec:forecast_evaluation}

Here we evaluate the predictive ability the proposed extension of the SEIR model where we assumed daily varying correlated random walk processes. Again, we compare predictive performance with comparable models from literature where independent-across-regions and piecewise constant with weekly or bi-weekly breakpoints random walk processes have been employed. 


To perform a full-fledged quantitative evaluation of the forecasts again we compare models through predictive performance, using the same measures as in Section 4.2.3 and, additionally, the ranked probability score and mean interval widths. The ranked probability score is useful for assessing that the forecasts are well calibrated and are calculated using the R package \texttt{scoringRules}. Mean interval widths are a measure of the accuracy of the forecasts - we want well calibrated forecasts that are maximally precise,calculated as the sample means of the obtained prediction intervals. 

\begin{figure}[H]
    \centering
    \includegraphics[scale=0.4]{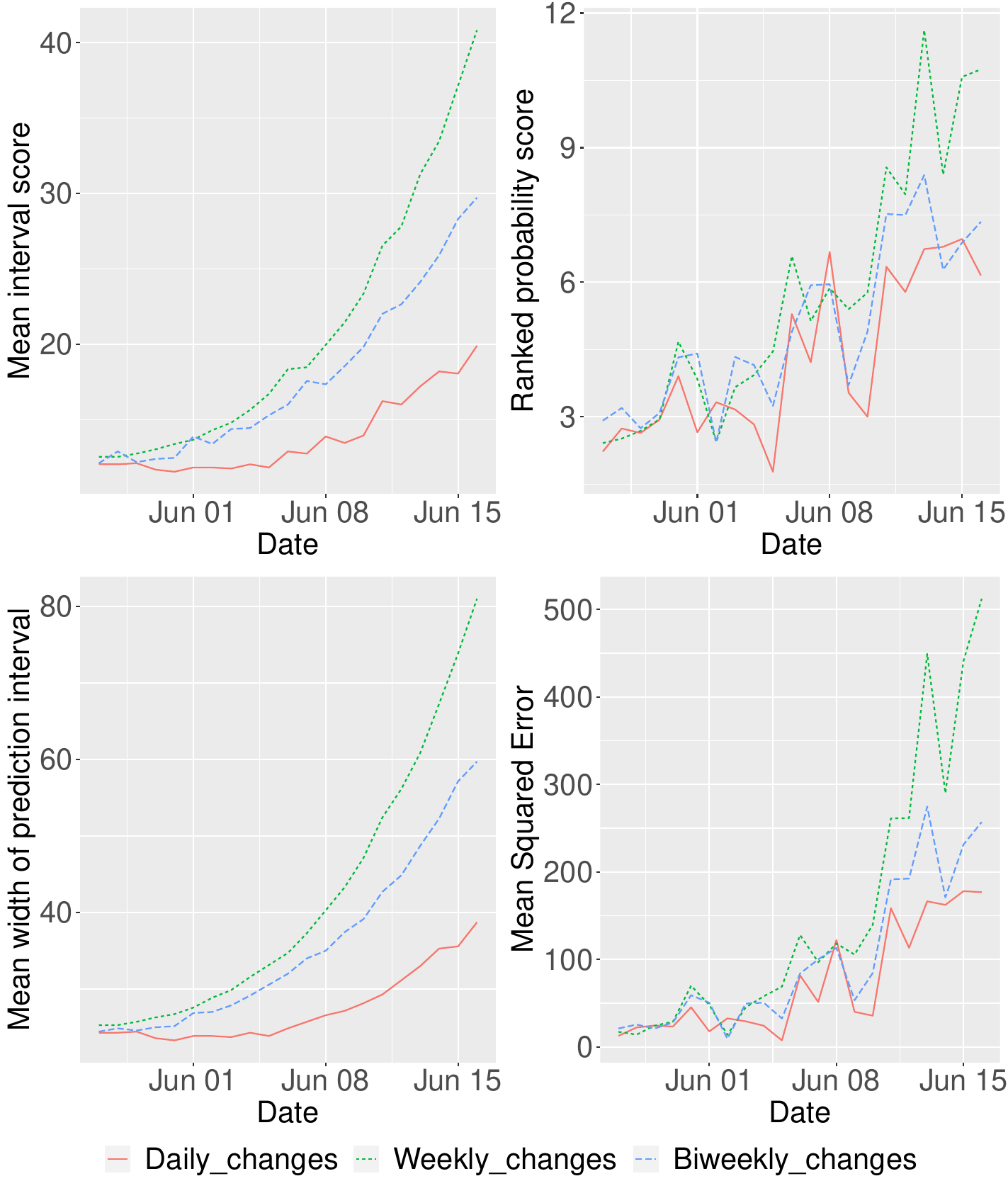}
    \caption{Mean interval scores (top left), ranked probability scores (top right) scores, mean widths (bottom left) of the $95\%$ prediction intervals and mean squared errors of mean forecasts (bottom right) calculated for each of the models under comparison by using data from 2020-02-16 to 2020-05-27 to predict the number of deaths for the period 2020-02-17 to 2020-06-16.}
    \label{fig:pred_comp}
\end{figure}

Figure \ref{fig:pred_comp}  presents mean values for the four different forecast evaluation metrics taken across the $m=7$ regions of England under each of the three different models. It is clear from the visual inspection of the Figure that the prediction intervals of the proposed model have lower interval and probability scores compared with the corresponding scores of the models with independent piecewise random walk processes. Thus, by following the proposed modelling approach we perform more accurate forecasts for future number of deaths than the competitive models. Moreover, the interval scores in combination with the width of the prediction intervals (presented in the bottom of Figure \ref{fig:pred_comp}) provide useful insights for the forecasting ability of the models under comparison. The lower interval scores of the proposed model are mainly achieved due to the narrower prediction intervals that are obtained. In fact, this is the main gain of replacing the independent piecewise random walk processes of the existing models with jointly modelled processes with daily changes since in the latter model the uncertainty regarding the dynamics of the disease transmission is estimated more accurately. By looking the mean squared errors of the forecasts conducted by using the three models we conclude that the proposed model not only produces prediction intervals with better properties than those of the existing methods but the delivered predictions are also more concentrated around the true values. One more interesting conclusion obtained from the visual inspection of Figure \ref{fig:pred_comp} is that the gain of using the proposed modelling approach for making future forecasts is increasing over time. The measures visualised by Figure \ref{fig:pred_comp} imply that the prediction intervals of the models with independent and piecewise random walk processes are becoming less accurate than those obtained by the proposed model when we increase the prediction horizon of interest.  
Finally, in Table \ref{tab:table1} we summarize the results that are presented by Figure \ref{fig:pred_comp} by providing averages, over
time, of the measures that we used to assess the predictions that were obtained from the models under comparison. The numbers in Table \ref{tab:table1} indicate that modelling jointly the region-specific random walk process and allowing for daily changes in the states of the process we obtain more accurate forecasts for future deaths than those obtained from the competitive models.

\begin{table}[H]
\caption{\label{tab:table1}Predictive performance of the different models: for each region the average predictive measure for the prediction period from 2020-02-17 to 2020-06-16 is reported.}
\scalebox{0.8}{
 \fbox{%
\begin{tabular}{l||l|c||c|c}
&Region &Daily changes& Weekly changes& Bi-weekly changes\\
\hline
\multirow{4}{*}{\parbox{5cm}{Mean interval score}}
&South West&$7.59$&$8$&$5$\\
&South East&$16.07$&$20$&$21$\\
&London &$6.30$&$20$&$9$  \\
&East of England &$19.02$&$22$&$20$  \\
&Midlands &$15.68$&$25$&$22$  \\
&North West &$14.32$&$27$&$22$  \\
&North East and Yorkshire &$18.02$&$24$&$25$  \\
\hline
\multirow{4}{*}{\parbox{5cm}{Ranked probability score}}
&South West&$1.94$&$2.16$&$3.89$\\
&South East&$5.84$&$3.95$&$5.24$\\
&London &$3.04$&$6.99$&$2.64$  \\
&East of England &$8.70$&$7.43$&$6.46$  \\
&Midlands &$3.34$&$4.29$&$4.14$  \\
&North West &$3.18$&$8.88$&$4.86$  \\
&North East and Yorkshire &$3.85$&$6.36$&$7.48$  \\
\hline
\multirow{4}{*}{\parbox{5cm}{Mean width of prediction intervals}}
&South West&$15.33$&$15.95$&$7.81$\\
&South East&$31.19$&$40.57$&$43.38$\\
&London &$12.52$&$39.14$&$18.38$  \\
&East of England &$35.62$&$44.76$&$40.00$  \\
&Midlands &$31.67$&$49.52$&$44.10$  \\
&North West &$28.90$&$53.67$&$44.62$  \\
&North East and Yorkshire &$36.38$&$47.90$&$51.19$  \\
\hline
\multirow{4}{*}{\parbox{5cm}{Mean squared error}}
&South West&$12.08$&$16.10$&$30.78$\\
&South East&$118.00$&$69.98$&$119.06$\\
&London &$22.69$&$214.00$&$17.90$  \\
&East of England &$234.04$&$211.44$&$154.32$  \\
&Midlands &$32.83$&$64.85$&$62.11$  \\
&North West &$29.91$&$323.14$&$103.86$  \\
&North East and Yorkshire &$59.15$&$164.48$&$212.21$  \\
\end{tabular}
}}
\end{table}

\section{Discussion}
\label{sec:conclusions}

We have proposed a modelling approach that realistically extends a SEIR model to account for environmental stochasticity and enables computationally efficient Bayesian inference, addressing the methodological limitations emerged in recent work on Covid-19 in England \citep{flaxman2020estimating,birrell2021real} and other European countries \citep{bouranis2022bayesian}. 
In particular, we have introduced a GMRF to model jointly the evolution of the disease transmission process over time and across regions. The proposed approach results in a model with increasing dimension over time. To tackle the increasing dimensionality and keep the inference timely, we have also developed an MCMC algorithm to draw samples from the posterior of interest by combining existing techniques that deal with high-dimensional, heavy-tailed and highly correlated targets. 

The proposed model has improved predictive performance over existing models that assume piece-wise constant and independent stochastic transmission processes, failing to appropriately quantify the uncertainty in predictions of disease sequelae.  In contrast, the 
prediction intervals resulting from the proposed modelling approach are shown to be well calibrated (see Figure \ref{fig:interval_scores_A}).

Additionally, our MCMC algorithm is designed to directly tackle a number of challenges: (i) the lack of identifiability of latent states and model parameters resulting from indirect and noisy observations leading to diffuse, heavy-tailed, posterior distributions; (ii) the computational cost stemming from the iterative process of solving the ODE system for likelihood evaluations and the high dimensionality of the latent space, especially under fine temporal discretisation, which can severely affect MCMC mixing and convergence. Our proposed framework  addresses the above issues in an efficient and scalable manner, expanding the scope of Bayesian inference for this class of models.

There are various aspects that should be explored in future work. Firstly, although our MCMC algorithm outperforms standard methods in terms of mixing and ESS, further enhancements could broaden its applicability and efficiency. More specifically, tempering ideas \citep{tawn2020weight} could be explored to sample the static parameters and deal with the model misspecification issues \citep{grunwald2017inconsistency} that often arise in epidemic inference; secondly, gradient-based methods (see {\it e.g.} \cite{titsias2018auxiliary}) could be investigated to improve sampling of the dynamic part of the model.

However, the use of tempering techniques requires extra care as they cannot be easily combined with our approach to sampling the high-dimensional dynamic states. If not carefully embedded, they may lead to substantial increase in the overall computational cost, especially when using a fine temporal resolution or large latent state spaces. For the use of gradient-based approaches, it is important to note that computing the gradients of the likelihood might also involve additional computational costs, jeopardizing the capacity for timely inference, a key requirement during any pandemic. Gaussian process-based emulation and calibration techniques (see {\it e.g.} \cite{farah2014bayesian} and \cite{seymour2022bayesian}) could be considered to minimize these costs in the context of the proposed modelling approach. Finally, an interesting research direction appears to be the approximation of the model by borrowing ideas from the machine learning and variational inference literature; see for example \cite{surjanovic2022parallel} for general methodology and \cite{ghosh2022differentiable} for applications in epidemics.  

\section{Code and data}

Replication code for the simulation study and data generation are available at \url{https://gitlab.com/aggelisalexopoulos/seir_gmrf}. 
For access to real data please contact the corresponding author.

\bibliographystyle{Chicago}
\bibliography{refs}

@article{birrell2011bayesian,
  title={Bayesian modeling to unmask and predict influenza A/H1N1pdm dynamics in London},
  author={Birrell, Paul J and Ketsetzis, Georgios and Gay, Nigel J and Cooper, Ben S and Presanis, Anne M and Harris, Ross J and Charlett, Andr{\'e} and Zhang, Xu-Sheng and White, Peter J and Pebody, Richard G and others},
  journal={Proceedings of the National Academy of Sciences},
  volume={108},
  number={45},
  pages={18238--18243},
  year={2011},
  publisher={National Academy Sciences}
}

@article{birrell2024real,
  title={Real-time modelling of the SARS-CoV-2 pandemic in England 2020-2023: A challenging data integration},
  author={Birrell, Paul J and Blake, Joshua and Kandiah, Joel and Alexopoulos, Angelos and van Leeuwen, Edwin and Pouwels, Koen and Ghosh, Sanmitra and Starr, Colin and Walker, Ann Sarah and House, Thomas A and others},
  journal={arXiv preprint arXiv:2408.04178},
  year={2024}
}

@article{rosenthal2011optimal,
  title={Optimal proposal distributions and adaptive MCMC},
  author={Rosenthal, Jeffrey S and others},
  journal={Handbook of Markov Chain Monte Carlo},
  volume={4},
  number={10.1201},
  year={2011},
  publisher={Chapman \& Hall/CRC Boca Raton, FL}
}

@article{surjanovic2022parallel,
  title={Parallel Tempering With a Variational Reference},
  author={Surjanovic, Nikola and Syed, Saifuddin and Bouchard-C{\^o}t{\'e}, Alexandre and Campbell, Trevor},
  journal={arXiv preprint arXiv:2206.00080},
  year={2022}
}

@article{seymour2022bayesian,
  title={Bayesian nonparametric inference for heterogeneously mixing infectious disease models},
  author={Seymour, Rowland G and Kypraios, Theodore and O’Neill, Philip D},
  journal={Proceedings of the National Academy of Sciences},
  volume={119},
  number={10},
  pages={e2118425119},
  year={2022},
  publisher={National Acad Sciences}
}

@article{farah2014bayesian,
  title={Bayesian emulation and calibration of a dynamic epidemic model for A/H1N1 influenza},
  author={Farah, Marian and Birrell, Paul and Conti, Stefano and Angelis, Daniela De},
  journal={Journal of the American Statistical Association},
  volume={109},
  number={508},
  pages={1398--1411},
  year={2014},
  publisher={Taylor \& Francis}
}

@article{grunwald2017inconsistency,
  title={Inconsistency of Bayesian inference for misspecified linear models, and a proposal for repairing it},
  author={Gr{\"u}nwald, Peter and Van Ommen, Thijs},
  journal={Bayesian analysis},
  volume={4},
  number={12},
  pages={1069-1103},
  year={2017}
}

@article{tawn2020weight,
  title={Weight-preserving simulated tempering},
  author={Tawn, Nicholas G and Roberts, Gareth O and Rosenthal, Jeffrey S},
  journal={Statistics and Computing},
  volume={30},
  number={1},
  pages={27--41},
  year={2020},
  publisher={Springer}
}

@article{valderrama2019mcmc,
  title={MCMC techniques for parameter estimation of ODE based models in systems biology},
  author={Valderrama-Baham{\'o}ndez, Gloria I and Fr{\"o}hlich, Holger},
  journal={Frontiers in Applied Mathematics and Statistics},
  volume={5},
  pages={55},
  year={2019},
  publisher={Frontiers Media SA}
}

@article{chis2011structural,
  title={Structural identifiability of systems biology models: a critical comparison of methods},
  author={Chis, Oana-Teodora and Banga, Julio R and Balsa-Canto, Eva},
  journal={PloS one},
  volume={6},
  number={11},
  pages={e27755},
  year={2011},
  publisher={Public Library of Science San Francisco, USA}
}

@article{flaxman2020estimating,
  title={Estimating the effects of non-pharmaceutical interventions on COVID-19 in Europe},
  author={Flaxman, Seth and Mishra, Swapnil and Gandy, Axel and Unwin, H Juliette T and Mellan, Thomas A and Coupland, Helen and Whittaker, Charles and Zhu, Harrison and Berah, Tresnia and Eaton, Jeffrey W and others},
  journal={Nature},
  volume={584},
  number={7820},
  pages={257--261},
  year={2020},
  publisher={Nature Publishing Group}
}

@article{jordan2017evaluating,
  title={Evaluating probabilistic forecasts with scoringRules},
  author={Jordan, Alexander and Kr{\"u}ger, Fabian and Lerch, Sebastian},
  journal={arXiv preprint arXiv:1709.04743},
  year={2017}
}

@article{grenfell1992chance,
  title={Chance and chaos in measles dynamics},
  author={Grenfell, BT},
  journal={Journal of the Royal Statistical Society: Series B (Methodological)},
  volume={54},
  number={2},
  pages={383--398},
  year={1992},
  publisher={Wiley Online Library}
}

@article{jacquez1996compartmental,
  title={Compartmental analysis in biology and medicine, BioMedware},
  author={Jacquez, John A},
  journal={Ann Arbor, MI},
  volume={512},
  year={1996}
}

@article{cazelles2018accounting,
  title={Accounting for non-stationarity in epidemiology by embedding time-varying parameters in stochastic models},
  author={Cazelles, Bernard and Champagne, Clara and Dureau, Joseph},
  journal={PLoS computational biology},
  volume={14},
  number={8},
  pages={e1006211},
  year={2018},
  publisher={Public Library of Science San Francisco, CA USA}
}

@inproceedings{ghosh2022differentiable,
  title={Differentiable Bayesian inference of SDE parameters using a pathwise series expansion of Brownian motion},
  author={Ghosh, Sanmitra and Birrell, Paul J and De Angelis, Daniela},
  booktitle={International Conference on Artificial Intelligence and Statistics},
  pages={10982--10998},
  year={2022},
  organization={PMLR}
}

@article{bailey1971estimation,
  title={The estimation of parameters from population data on the general stochastic epidemic},
  author={Bailey, Norman TJ and Thomas, Anthony S},
  journal={Theoretical Population Biology},
  volume={2},
  number={3},
  pages={253--270},
  year={1971},
  publisher={Elsevier}
}

@article{lekone2006statistical,
  title={Statistical inference in a stochastic epidemic SEIR model with control intervention: Ebola as a case study},
  author={Lekone, Phenyo E and Finkenst{\"a}dt, B{\"a}rbel F},
  journal={Biometrics},
  volume={62},
  number={4},
  pages={1170--1177},
  year={2006},
  publisher={Wiley Online Library}
}

@article{dukic2012tracking,
  title={Tracking epidemics with Google flu trends data and a state-space SEIR model},
  author={Dukic, Vanja and Lopes, Hedibert F and Polson, Nicholas G},
  journal={Journal of the American Statistical Association},
  volume={107},
  number={500},
  pages={1410--1426},
  year={2012},
  publisher={Taylor \& Francis}
}

@article{keeling2002understanding,
  title={Understanding the persistence of measles: reconciling theory, simulation and observation},
  author={Keeling, Matt J and Grenfell, Bryan T},
  journal={Proceedings of the Royal Society of London. Series B: Biological Sciences},
  volume={269},
  number={1489},
  pages={335--343},
  year={2002},
  publisher={The Royal Society}
}

@book{anderson1992infectious,
  title={Infectious diseases of humans: dynamics and control},
  author={Anderson, Roy M and May, Robert M},
  year={1992},
  publisher={Oxford university press}
}

@article{dureau2013capturing,
  title={Capturing the time-varying drivers of an epidemic using stochastic dynamical systems},
  author={Dureau, Joseph and Kalogeropoulos, Konstantinos and Baguelin, Marc},
  journal={Biostatistics},
  volume={14},
  number={3},
  pages={541--555},
  year={2013},
  publisher={Oxford University Press}
}

@article{andrieu2008tutorial,
  title={A tutorial on adaptive MCMC},
  author={Andrieu, Christophe and Thoms, Johannes},
  journal={Statistics and computing},
  volume={18},
  pages={343--373},
  year={2008},
  publisher={Springer}
}

@article{vLeS20,
author = {van Leeuwen, Edwin and {{PHE Joint modelling cell}} and Sandmann, Frank},
doi = {10.1101/2020.06.03.20067793},
journal = {medRxiv},
month = {jun},
pages = {2020.06.03.20067793},
publisher = {Cold Spring Harbor Laboratory Press},
title = {{Augmenting contact matrices with time-use data for fine-grained intervention modelling of disease dynamics: A modelling analysis}},
year = {2020}
}

@article{girolami2011riemann,
  title={Riemann manifold {L}angevin and {H}amiltonian {M}onte {C}arlo methods},
  author={Girolami, Mark and Calderhead, Ben},
  journal={Journal of the Royal Statistical Society: Series B (Statistical Methodology)},
  volume={73},
  number={2},
  pages={123--214},
  year={2011},
  publisher={Wiley Online Library}
}

@article{titsias2018auxiliary,
  title={Auxiliary gradient-based sampling algorithms},
  author={Titsias, Michalis K and Papaspiliopoulos, Omiros},
  journal={Journal of the Royal Statistical Society. Series B (Statistical Methodology)},
  volume={80},
  number={4},
  pages={749--767},
  year={2018},
  publisher={JSTOR}
}

@article{beskos2008mcmc,
  title={{MCMC} methods for diffusion bridges},
  author={Beskos, Alexandros and Roberts, Gareth and Stuart, Andrew and Voss, Jochen},
  journal={Stochastics and Dynamics},
  volume={8},
  number={03},
  pages={319--350},
  year={2008},
  publisher={World Scientific}
}

@article{neal1998regression,
  title={Regression and classification using Gaussian process priors},
  author={Neal, M. Radford},
  journal={Bayesian statistics (Alcoceber, 1998)},
  volume={6},
  pages={475},
  year={1998}
}

@article{chib2010tailored,
  title={Tailored randomized block {MCMC} methods with application to {DSGE} models},
  author={Chib, Siddhartha and Ramamurthy, Srikanth},
  journal={Journal of Econometrics},
  volume={155},
  number={1},
  pages={19--38},
  year={2010},
  publisher={Elsevier}
}

@book{rue2005gaussian,
  title={Gaussian Markov random fields: theory and applications},
  author={Rue, Havard and Held, Leonhard},
  year={2005},
  publisher={Chapman and Hall/CRC}
}

@article{knock2021key,
  title={Key epidemiological drivers and impact of interventions in the 2020 {SARS-CoV-2} epidemic in {England}},
  author={Knock, Edward S and Whittles, Lilith K and Lees, John A and Perez-Guzman, Pablo N and Verity, Robert and FitzJohn, Richard G and Gaythorpe, Katy AM and Imai, Natsuko and Hinsley, Wes and Okell, Lucy C and others},
  journal={Science Translational Medicine},
  volume={13},
  number={602},
  pages={eabg4262},
  year={2021},
  publisher={American Association for the Advancement of Science}
}

@article{bouranis2022bayesian,
  title={Bayesian analysis of diffusion-driven multi-type epidemic models with application to {COVID-19}},
  author={Bouranis, Lampros and Demiris, Nikolaos and Kalogeropoulos, Konstantinos and Ntzoufras, Ioannis},
  journal={arXiv preprint arXiv:2211.15229},
  year={2022}
}

@article{birrell2021real,
  title={Real-time nowcasting and forecasting of {COVID-19} dynamics in {E}ngland: the first wave},
  author={Birrell, Paul and Blake, Joshua and Van Leeuwen, Edwin and Gent, Nick and De Angelis, Daniela},
  journal={Philosophical Transactions of the Royal Society B},
  volume={376},
  number={1829},
  pages={20200279},
  year={2021},
  publisher={The Royal Society}
}

@article{haario2001adaptive,
  title={An adaptive Metropolis algorithm},
  author={Haario, Heikki and Saksman, Eero and Tamminen, Johanna},
  journal={Bernoulli},
  pages={223--242},
  year={2001},
  publisher={JSTOR}
}

\end{document}